\documentclass{jpconf}

\usepackage{hyperref,graphicx,subfigure}
 \hypersetup{colorlinks=true, linkcolor=black, urlcolor=black, citecolor=black, bookmarksopen}
 
\usepackage{amssymb, amsmath}
\usepackage[square,sort&compress,merge]{natbib}
\setcitestyle{numbers}
 
\newcommand{\be}[0]{\begin{equation}}
\newcommand{\ee}[0]{\end{equation}}
\newcommand{\ba}[0]{\begin{eqnarray}}
\newcommand{\ea}[0]{\end{eqnarray}}
\newcommand{\Z}[0]{\mathbb{Z}}
\newcommand{\q}[0]{\quad}

\begin{document}

\title{Towards computational insights into the large-scale structure of spin foams}

\author{Bianca Dittrich and Frank C Eckert}

\address{MPI f. Gravitational Physics, Albert Einstein Institute, \\ Am M\"uhlenberg 1, D-14476 Potsdam, Germany}

\ead{dittrich@aei.mpg.de, frank.eckert@aei.mpg.de}

\begin{abstract}

Understanding the large-scale physics is crucial for the spin foam approach to quantum gravity. We tackle this challenge from a statistical physics perspective using simplified, yet feature-rich models. In particular, this allows us to explicitly answer whether broken symmetries will be restored by renormalization: We observe a weak phase transition in both Migdal-Kadanoff and tensor network renormalization. In this work we give a concise presentation of the concepts, results and promises of this new direction of research.

\end{abstract}

\section{Will broken symmetries be restored in the large-scale limit?}

The investigation of the large-scale structure of spin foam models is crucial.
There are three reasons for this.

First, spin foam models can be understood as giving a meaning to the path integral for quantum gravity 
\citep{Perez:2003vx,*Alejandro,Bahr:2011yc}
and, in order to do so in a well defined way, spin foams introduce a regularization. 
The regularization is based on a discretization by means of a lattice, triangulation or two-complex and this will generically  break (global) Lorentz and (local) diffeomorphism invariance \cite{Dittrich:2008pw,*Bahr:2009ku,*Bahr:2009mc}. Will these symmetries be restored in the large-scale limit\footnote{
The auxiliary discretization and its large-scale limit is not to be confused with another source of discreteness (Planck scale discreteness) in the theory \cite{Rovelli:1994ge,*Ashtekar:1996eg,*Dittrich:2007th,*Rovelli:2007ep}, namely in the spectra of observables, and its corresponding $\hbar \rightarrow 0$ limit. The relationship between the two is an open issue.} \cite{Bahr:2009qc,*Bahr:2010cq,*Bahr:2011uj}?
As diffeomorphisms constitute the fundamental symmetry of general relativity this question is essential to demonstrate the viability of spin foams as a quantum theory of gravity.

Second, it is not clear whether such a statistical limit will give a space-time modelled by a smooth four-dimensional manifold. It is imperative, however, to realize a smooth continuum limit in order to
validate any approach to quantum gravity and to make predictions of measurable physical effects.
Studies in (causal) dynamical triangulations have shown \cite{Ambjorn:1995jt,*Ambjorn:2005jj,*Ambjorn:2011ph} that degenerate configurations may appear and are relevant.
It is therefore necessary to understand whether, when and how pathological structures appear and dominate the state sum. In other words, it is necessary to understand the phase structure of spin foams. 

Third, even though spin foams currently enjoy much attention little progress has been made in probing the large-scale limit.
The question is important and other approaches to quantum gravity have progressed in this direction. Therefore, understanding the large-scale structure of spin foams is now one of the most pressing problems in the field.

The reason why our understanding of the large-scale physics of spin foams is stagnating is simple: the models are very complex and mathematically involved, making it difficult to perform even simple calculations, let alone to use the computer as a computational tool.

We approach this issue by radically simplifying the mature spin foam models. 
This will 
facilitate the development of renormalization tools, methods and most importantly numerical 
algorithms. Furthermore, by studying the renormalization behavior of increasingly more 
complicated models, it is possible to deduce properties for the full models. In this work we 
start with the study of the simplest models:
 Instead of non-Abelian Lie groups we use finite Abelian groups. We also confine our models to regular lattices (instead of general two-complexes), which allows to apply  effective approximation schemes. As is argued in \cite{Bahr:2009qc,*Bahr:2010cq,*Bahr:2011uj} a restoration of 
diffeomorphism symmetry should also ensure independence from the chosen underlying lattice 
or triangulation. Hence the restriction to regular lattices is justified. Finally, we consider 
dimensionally reduced versions of the four-dimensional models, so called \emph{spin net models}, which in fact will be the focus of this article.  Interestingly, spin net models in two dimensions share 
many statistical properties with spin foam models in four dimensions \cite{Ito:1985bv}.

Simplification as put forward here is not only done so that the resulting models become computationally manageable. Rather, we choose to simplify in such a way that important features are conserved and meaningful problems can be studied. In particular, important (gauge) symmetries are preserved, the model space includes a topological ($BF$ like) theory and we introduce certain \emph{Abelian cutoff models}, which break the symmetries in a controlled way and can be used to study the question of whether broken symmetries are being restored under renormalization.

In doing so we connect with statistical and condensed matter physics. We can make use of the power of computers and of highly developed methods, which we refine and adapt to suit the questions we are asking. Ultimately, this enables us do calculations, to study the large-scale physics of simplified spin foam and spin net models and to explicitly answer the question of symmetry restoration. While interesting in itself, we see promise in this as a first step towards computational insights into the large-scale structure of mature spin foam models. Furthermore, as we will see, results obtained across a range of finite groups can inform about the analogue and more complicated situation with the corresponding continuous group.

This article is a concise presentation and discussion of work published in \cite{dem} to which we refer the reader for details. The outline is as follows. We first construct a setting in which we have simple models and a gauge symmetry which can be broken, simple enough to analyze it computationally. We then discuss two numerical renormalization schemes, Migdal-Kadanoff and tensor network renormalization, and the weak phase transition we observe in both cases. Finally, we outline the next steps to be done.

\section{Spin net models: controlled simplification retaining symmetries}

Spin net models are dimensionally reduced versions of spin foam models \cite{Bahr:2011yc,dem}. They are a class  of lattice theories that include the Ising model. In this article we will solely focus on simple 2D spin net models with finite Abelian groups. A more general and detailed treatment can be found in \cite{dem}.

Given a regular, oriented graph and an element $g$ of the finite Abelian group $\Z_q$ a spin net model is defined via its partition function
\be\label{Z-hol1}
 Z=\frac1{q^{\sharp v}}\sum_{\{g_v\}}\prod_e w(g_{\raisebox{-1.25pt}{$\scriptstyle s_e$}}g^{-1}_{t_e}),
\ee
where $\sharp v$ denotes the number of vertices in the graph, $\sum_{\{g_v\}} = \prod_v \sum_{g_v}$ is the sum over all assignments of group elements to the graph's vertices and $\prod_e$ is the product over all edges.
Here, the index $s_e$ (resp. $t_e$) gives the source (resp. target) vertex and $w$ is the weight function.
In analogy to spin foam models we call this way of writing the state sum the \emph{holonomy representation}.

The \emph{spin net representation}, in contrast, is obtained by expressing the state sum in terms of irreducible representations of the group. This is done via a group Fourier transform which is given by the character decomposition,
\be\label{za3}
w(g) = \sum_{k=0}^{q-1}   \tilde w_k \, \chi_k(g)  \qquad
\tilde w_k = q^{-1} \sum_{g=0}^{q-1}  w(g)  \overline{\chi}_k(g)\, .
\ee
For $\Z_q$, the irreducible representations are one dimensional, they are labeled by $ k \in \Z_q$ and the characters, given by $\chi_k(g)=\exp(({2\pi i}/{q}) k\cdot g)$ for $g \in \Z_q$, 
obey $\chi_k(g_1\cdot g_2)=\chi_k(g_1) \cdot \chi_k( g_2)$ and $\chi_k(g^{-1})=\chi_k^{-1}(g)=\overline{\chi}_k(g)$. Furthermore, the weight function with trivial Fourier coefficients gives the $q$-periodic delta function over the group, $\delta^{(q)} = \sum_k \chi_k(g)$, satisfying $q^{-1}\sum_g\delta^{(q)}(g)f(g)=f(0)$. With this, the state sum in the spin net representation takes the form
\be \label{Z-sn-abelian}
Z= \sum_{\{k_e\}}   \prod_e \tilde w_{k_e}  \prod_v \tilde{P}^v(\{k_e\}_{e\supset v}),\qquad
\tilde{P}^v(\{k_e\}_{e \supset v}):=
\delta^{(q)} \!\big( \sum_{e \supset v} \varepsilon^e_v k_e\big),
\ee
where we have introduced an orientation coefficient $\varepsilon^e_v$ which equals $+1$ (resp. $-1$) if $v$ is $e$'s source (resp. target). 

In equation \eqref{Z-sn-abelian} the vertex projector $\tilde{P}^v$ imposes Gau\ss~constraints at each vertex. More mathematically, this corresponds to a projection onto the irreducible subrepresentation in the tensor product of (dual) representations associated to (incoming) outgoing edges. The projection can be restricted even further in order to construct more complicated models. We will exemplify one such case here.

A choice of $\tilde w(k) \equiv 1$ in the spin net representation (corresponding to $w(g) = \delta^{(q)}(g)$ in the holonomy representation) defines a topological model without local degrees of freedom, also called $BF$ theory.\footnote{ The name $BF$ theory origins from the spin foam models. There $BF$ theory models are distinguished by the appearance of translation symmetries, equation (\ref{transisn}), from other spin foam models. As these symmetries also appear in the corresponding spin net models, we will apply this name for these spin nets.}
 From a statistical physics perspective, it is a `frozen' model at zero temperature: the group elements of each connected component of the graph have to agree. This model is invariant under a gauge transformation, also known as \emph{translation symmetry}, given by
\be\label{transisn}
k_e \mapsto k'_e=k_e + \sum_{f \supset e} \varepsilon^f_e k_f,
\ee
where the sum is over faces bounding the respective edge with corresponding parameters $k_f \in \Z_q$. $\varepsilon^f_e$ is again an orientation factor. 

The implementation of a certain non-trivial projector then leads to \emph{Abelian cutoff models} which break this gauge symmetry. The name stems from the fact that the effect of the desired projector restriction can also be achieved by a restriction on weights $\tilde{w}(k)$, namely
\be\label{Aco}
\tilde w(k) =
\begin{cases}
&1,\quad\text{for}\; |k|\leq K   \\
&0,\quad\text{for}\;|k|>K
\end{cases} \qquad, -\frac{q}{2}<k\leq\frac{q}{2}
\ee
effectively \emph{cutting off} the sum that defines the delta. Defined in this way the Abelian cutoff models are also symmetric, $\tilde{w}(k) = \tilde{w}(-k) \, \forall k \in \Z_q$, and thus, as desired, independent of edge orientations and certain subdivisions  \cite{Bojowald:2009im,*Bahr:2010my}.

The setup we now have includes a $BF$ model whose gauge symmetry is an analogue of diffeomorphism symmetry found in 3D $BF$ with group $SO(3)$ and cutoff models which implement non-trivial projectors and thereby break the gauge symmetry. Yet the models are much simpler to handle and this sets the stage for an explicit analysis of the problem whether the broken symmetries will be restored under renormalization.

\section{Simple, non-refinable approximation: Migdal-Kadanoff}

In order to determine the behaviour of cutoff models under renormalization a coarse graining procedure has to be implemented. However, coarse graining in general leads to a growing number of (non-local) couplings at each step and this makes it mandatory to apply an approximation scheme.

\begin{figure}[htbp]
\begin{center}
 \includegraphics[width=0.5\textwidth]{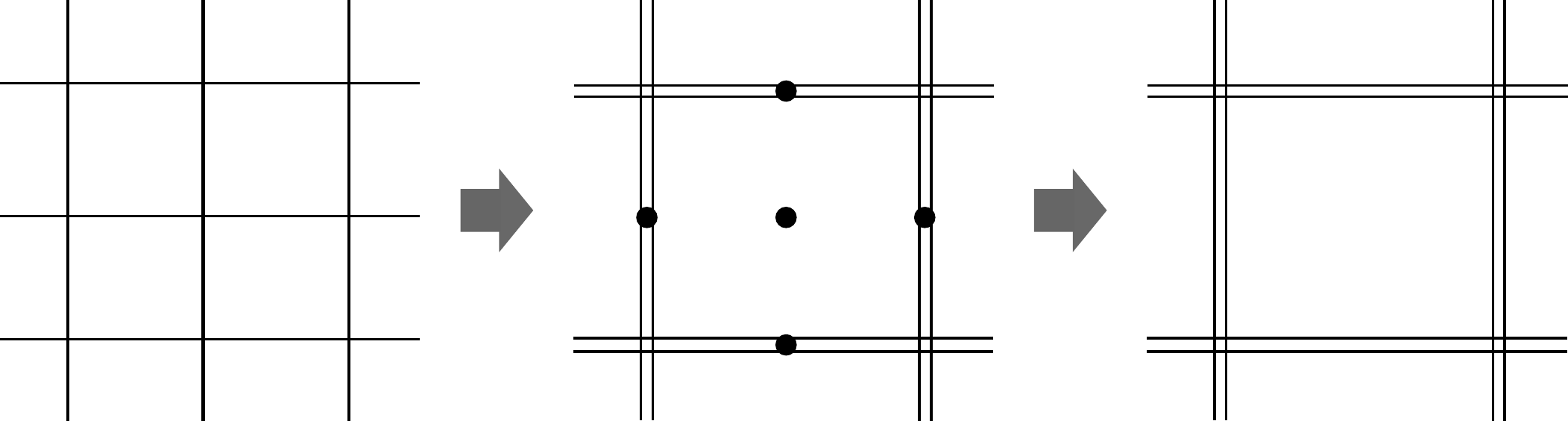} 
\caption{The derivation of MK recursion relations for 2D spin net models involves two steps: moving bonds and subsequently integrating out spins. \label{fig:mk}}
\end{center}
\end{figure}

The Migdal-Kadanoff (MK) approximation \cite{Migdal:1975zg, Kadanoff:1976jb} is a simple method to restrict the renormalized Hamiltonian to local couplings. This is done by moving bonds (links between vertices) away from vertices until the valency is effectively reduced to at most two, mirroring the one-dimensional situation in which no non-local couplings appear during coarse graining (see figure \ref{fig:mk}).

Kadanoff derived a lower bound on the free energy for this approximation \cite{Kadanoff:1976jb}. Further justification for this otherwise ad hoc method is drawn from fairly good predictions of fixed points and phase transitions \cite{Creutz:1982vz} and from comparison with Monte Carlo simulations \cite{Creutz:1979kf,*Creutz:1979zg}. The order of phase transitions and some fixed point of Kosterlitz-Thouless type \cite{Ito:1985bv}, however, are not correctly reproduced. The MK approximation is thus a simple and computationally efficient tool to probe the large-scale limit. It, however, does not offer means for systematic improvement, for instance by taking non--local couplings into account. 

The (isotropic) version of the MK recursion relation used here have the interesting property to coincide for 2D (Abelian) spin net models and 4D (Abelian) spin foam models \cite{Ito:1985bv}. Thus 2D spin nets can be seen as a statistical toy models for 4D spin foams.

\begin{figure}[htbp]
\begin{center}
 \mbox{
      \subfigure[fixed point structure for $\Z_4$]{\includegraphics[width=0.45\textwidth]{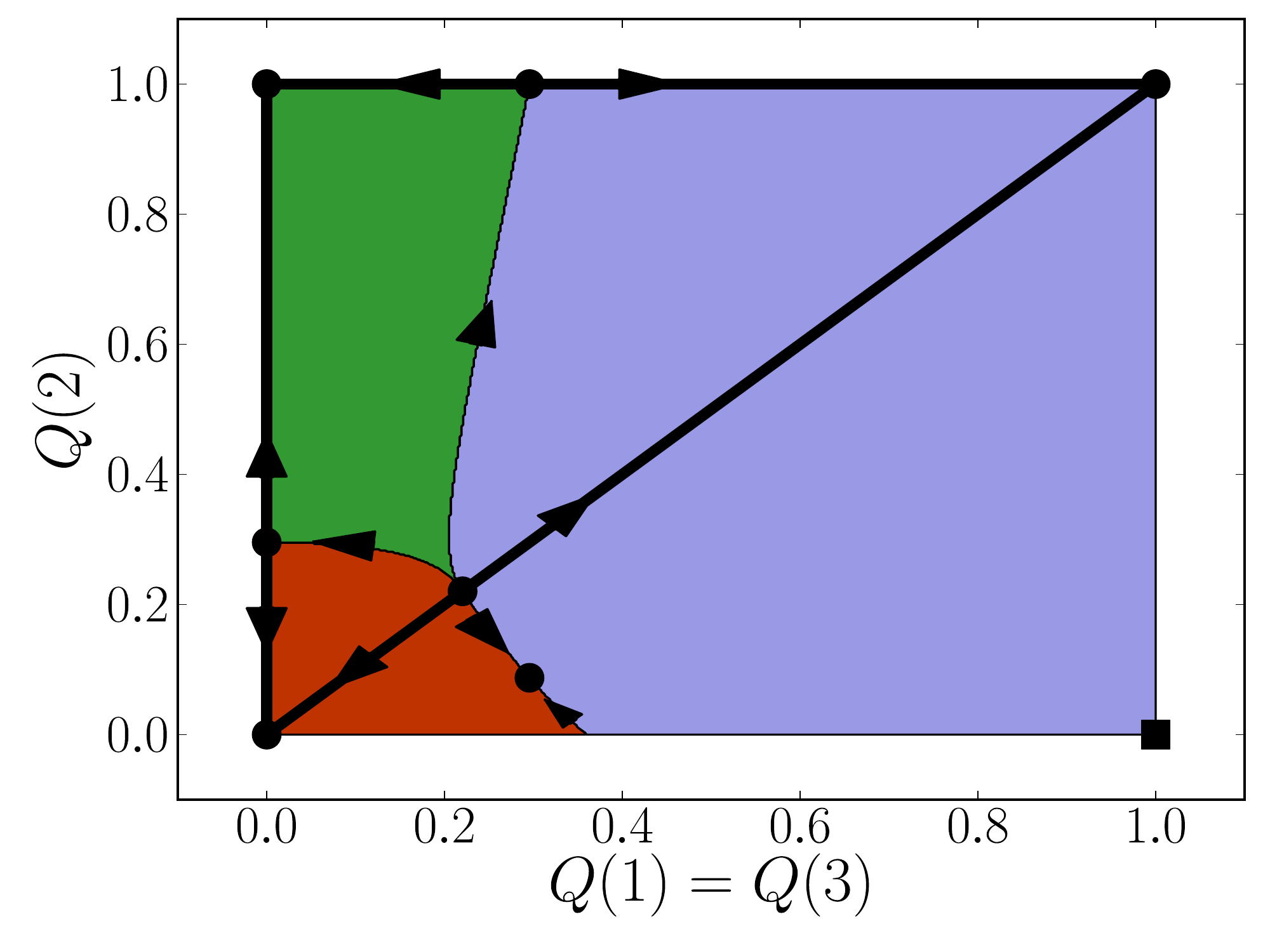}} \quad
      \subfigure[$q-K$ diagram showing weak phase transition]{\includegraphics[width=0.50\textwidth]{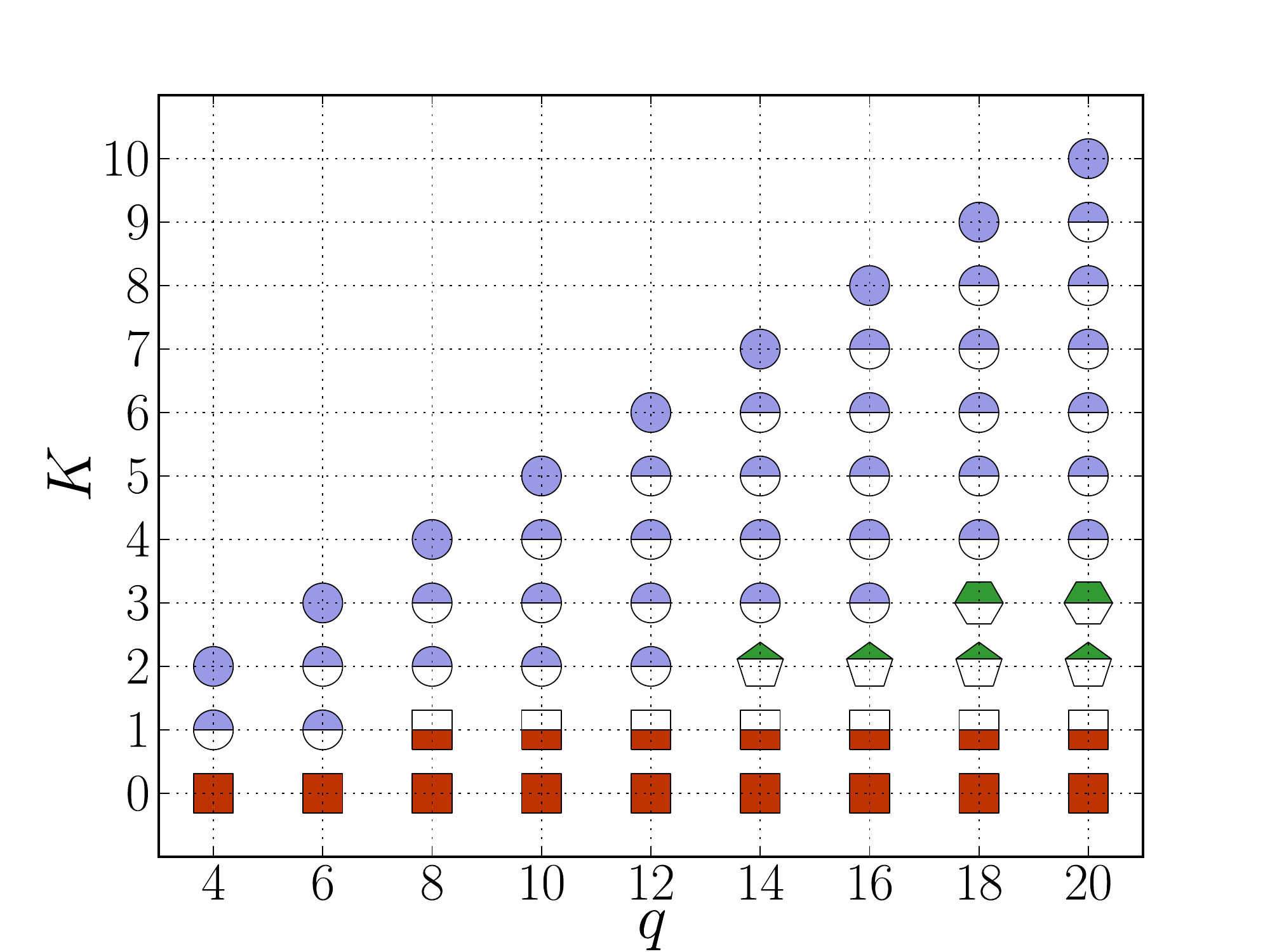}}
     }
\caption{(a) Parameter space of $\mathbb{Z}_4$ symmetric models and their renormalization flow. The three regions correspond to configurations flowing to HTF $(0,0)$, LTF / $BF$ $(1,1)$ and the `cyclic fixed point' $(0,1)$. (b) renormalization behaviour of cutoff models parametrized by $q$ and $K$. Markers half-filled at the top flow to BF/LTF (diagonal models, $K = q/2$), markers half-filled at the bottom flow to HTF (horizontal modesl with $K = 0$). Pentagonal and hexagonal markers denote unstable fixed points. \label{fig:mk-z4}}
\end{center}
\end{figure}

At the core of the MK approximation, the derivation of the recursion relations is based on the aforementioned bond-moving: In a systematic way, some terms (bonds) in the state sum are dropped while others are strengthened which leads to convolution of Fourier coefficients,
    \be
      w(g_{e'}) \rightarrow w^2(g_{e'}) = \sum_k  \left(  \sum_j \tilde{w}(k-j) \tilde{w}(j) \right) \chi_k(g_{e'}) .
    \ee
Convolution is a central feature of the recursion relations which, after normalization, $Q(k) = \tilde{w}(k) / \tilde{w}(0)$, take the form
    \be \label{mk-rr}
    Q_{n+1}(k) = \left( \frac{\sum_{j = 0}^{q-1} Q_{n}(k-j) Q_{n}(j)}{\sum_{j = 0}^{q-1} Q_{n}(-j) Q_{n}(j)} \right)^2
    \ee
and feature a rich fixed point structure exemplified in figure \ref{fig:mk-z4}a. 
From these relations it can be seen that the space of models they act on depends on $q$. We further restrict this space to symmetric models ($\tilde{w}(k) = \tilde{w}(-k) \, \forall k \in \Z_q$) which are preserved by \eqref{mk-rr}.
This includes the low and high temperature as well as the cutoff models. Due to small numerical errors, we have to artificially project onto symmetric configurations after each coarse graining step. The recursion relations feature two competing effects: convolution levels out differences in the Fourier coefficients and draws the model towards the low temperature fixed point  (LTF). The overall exponent, on the other hand, dampens all but $Q(0)$, drawing the configuration towards the high temperature fixed point  (HTF).

We now turn to the behaviour of cutoff models under MK renormalization. The results displayed in figure \ref{fig:mk-z4}b show the dependence of the result on $q$ and the phase transition between models flowing to LTF (or $BF$ phase) and those flowing to HTF. For most models, the dampening is too weak to stop them from flowing to $BF$, they do so quickly (5-15 iterations). Only a series of models in immediate proximity to HTF ($K=1,q \geq 8$) actually flow to HTF (after about 60 iterations). Thus we can conclude that the translation symmetry is restored for a range of cutoff models for finite $q$. 

However, the phase transition we observe is weak and becomes weaker with growing $q$. This can be seen in rising iteration numbers for fixed $K$ and growing $q$ (e.g.\ we observe $8,15,60$ iterations for $K=2$ and $q=8,10,12$ respectively). 
Starting with $q=14$ we also observe the appearance of quasi fixed points which are stable over large numbers of iterations.

These results are in accordance with similar considerations of MK relations for $q\rightarrow \infty$, i.e.\ for $U(1)$. The Mermin-Wagner theorem states that that the continuous symmetry cannot be spontaneously broken at finite temperature. There, however, is a Kosterlitz-Thouless phase transition of infinite order present, even though it is not detected by the MK approximation \cite{Ito:1985bv}.
Similarly, the generalized Ising fixed point between HTF and LTF is no longer present in the limit of large $q$ \cite{dem}.  

Similarly to the 2D models there is also a phase transition present in 4D $U(1)$ lattice gauge theory \cite{polya,*banks}. For 4D non--Abelian lattice gauge theories, however, the confinement conjecture states the absence of such a transition. All configurations would thus flow back to the HTF and in this case the translational symmetries would not be restored. Note that the HTF corresponds to a geometrically degenerate phase for the gravitational spin foam models. An essential question for future research will be whether this behavior is altered if non--trivial projectors are taken into account for the non--Abelian models. Such non--trivial projectors would no longer allow the model to be rewritten in standard lattice gauge theory form.

In 3D the gravitational model coincides with the $BF$ (i.e. LTF) phase, i.e. the projector in (\ref{Z-sn-abelian}) is the one on the maximal subspace of trivial representations. Here, for the finite Abelian group models  the tendency to flow to HTF is even stronger than in the 2D spin net  and 4D spin foam models. For $q\rightarrow\infty$ there is no phase transition within the MK framework \cite{Ito:1984ph}, which also holds for the non--Abelian Lie groups \cite{Muller:1984uz}. Hence a restoration of translational symmetries does not take place for the gravitational models 3D based on $SO(3)$. A question for future research will be, whether a phase transition takes place for the gravitational models with cosmological constant, which are based on quantum groups \cite{Turaev:1992hq}. These might resemble the finite group case more closely.

\section{Refinable, computationally intensive renormalization with tensor networks}

The tensor network renormalization (TNR) scheme \cite{levin,*Gu:2009dr} 
is more general than the MK approximation discussed so far. It is also refinable. However, it has so far only been successfully applied to 2D models.

Spin net models as in \eqref{Z-sn-abelian} can easily be rewritten in \emph{tensor network representation} which takes the general form of a tensor trace over tensors attached to vertices with indices contracted along edges,
\ba\label{au1}
Z=\sum_{a,b,c,d\ldots} T^{abcd} T^{aefg} T^{bhij} \cdots  \quad .
\ea
In order to arrive at this form it is only necessary to split the Fourier weights in \eqref{Z-sn-abelian} attached to the edges in two, $u(a) := \sqrt{\tilde{w}(a)}$, and to pull each half into the definition of the tensor at the vertex
\be
T^{abcd}= u(a) u(b) u(c) u(d) \tilde{P}(\{a,b,c,d\}) = u(a) u(b) u(c) u(d) \delta^{(q)}(a + b - c -d) .
\ee

The naive approach to coarse graining or blocking such a system will collect several tensors into one new tensor with several edges pointing in each direction or equivalently, with one edge with an exponentially growing index range. Here, the approximation comes in: it limits the index range appropriately. The renormalization group flow then depends on the choice of cutoff $D_c$.  This is in contrast to the MK method where the cutoff would be determined by the size of the group $\Z_q$. Indeed we find that in the TNR method the flow of Abelian cutoff models coincides for different choices of Abelian groups $\Z_q$, as long as the relation  $q\geq 4K+2$ holds.  
The cutoff  $D_c$ can be increased in order to refine the accuracy of predictions. In the case of the Ising model, this can indeed be shown explicitly, i.e.\ the accuracy of the critical temperatures found in the different approximations determined by $D_c$ increases with the cutoff $D_c$ \cite{dem}.

\begin{figure}
\begin{center}
 \mbox{
      \subfigure[square lattice]{\includegraphics[width=0.45\textwidth]{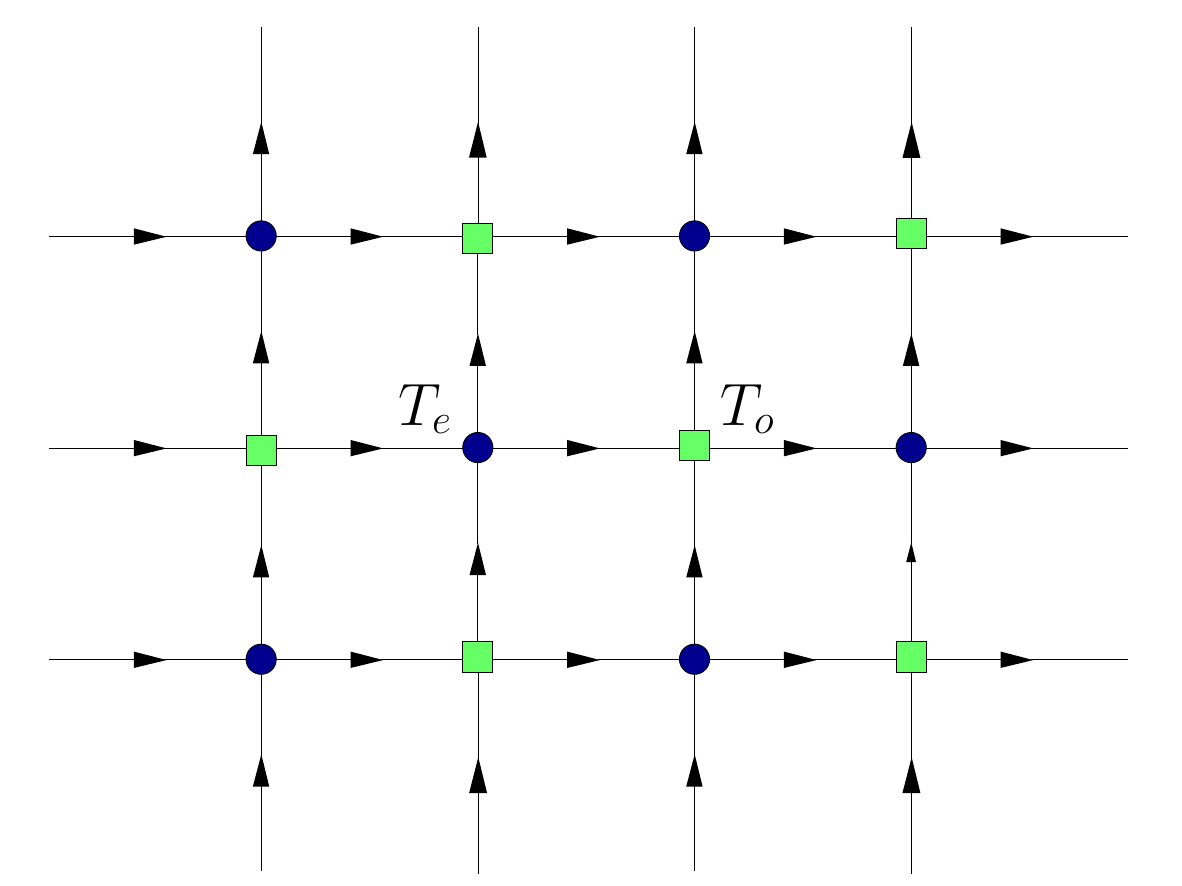}} \qquad
      \subfigure[splitting of vertices]{\includegraphics[width=0.45\textwidth]{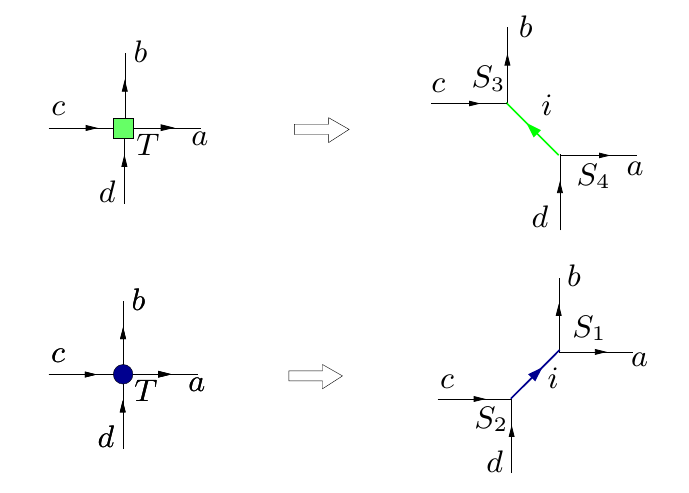}}
      }
\vspace*{0.2cm}
\caption{The square lattice with even (blue circles) and odd (green squares)  vertices.
\label{lattice}
}
\end{center}
\end{figure}

\begin{figure}
\begin{center}
 \mbox{
      \subfigure[contraction]{\includegraphics[width=0.60\textwidth]{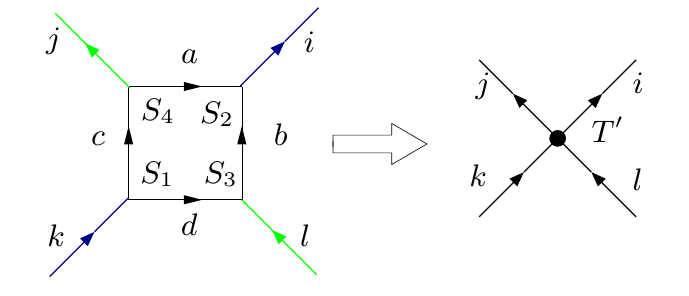}} \qquad
      \subfigure[coarse grained lattice]{\includegraphics[width=0.24\textwidth]{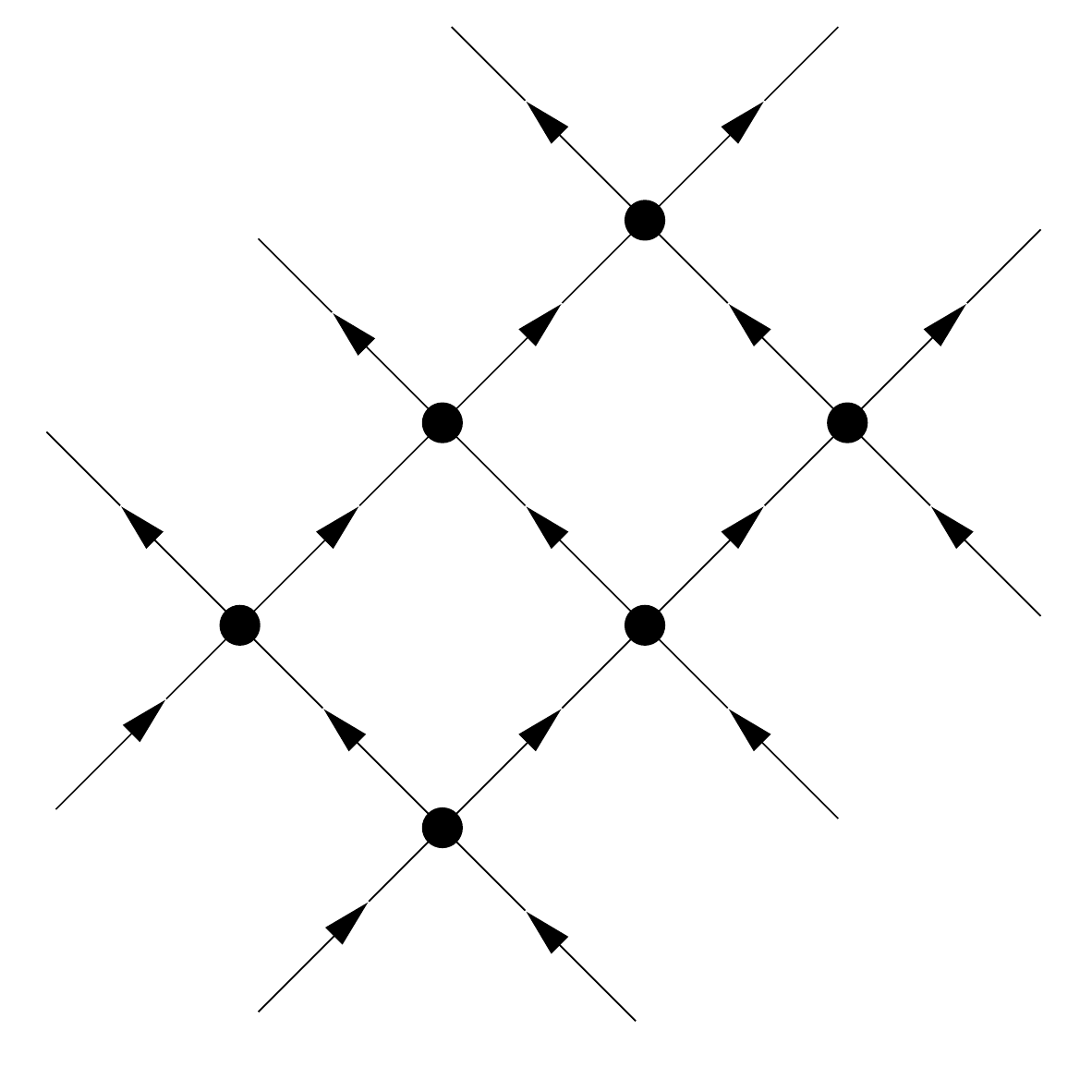}}
      }

\caption{$(a)$ Contraction of the four $S$ tensor to the new $T'$ tensor. $(b)$ The coarse grained lattice.
\label{fig:cvertex}
}
\end{center}
\end{figure}

The TNR method we follow here takes the state sum and reinterprets it as trace over matrices. As such, the state sum implicitly contains a sum over singular values of these matrices via a singular value decomposition. The approximation then proceeds by cutting off the sum over singular values, including only the $D_c$ largest among them:
\ba\label{t3}
T^{abcd} = M_1^{ab,cd} = \sum_{i=0}^{q^2-1} U_1^{ab,i} \lambda_i (V_1^\dagger)^{i,cd} \approx \sum_{i=0}^{D_c-1} (U_1^{ab,i} \sqrt{\lambda_i}) (\sqrt{\lambda_i} (V_1^\dagger)^{i,cd}) = \sum_i S_1^{ab,i} S_2^{cd,i} \\
T^{abcd} = M_1^{cb,ad} = \sum_{i=0}^{q^2-1} U_1^{cb,i} \lambda_i (V_1^\dagger)^{i,ad} \approx \sum_{i=0}^{D_c-1} (U_1^{cb,i} \sqrt{\lambda_i}) (\sqrt{\lambda_i} (V_1^\dagger)^{i,ad}) = \sum_i S_3^{cb,i} S_4^{ad,i}
\ea
The blocking can then be performed by contracting four tensors, 
\ba
 {T'}^{ijkl}=\sum_{a,b,c,d} S_2^{ab,i}S_4^{ac,j} S_1^{dc,k}S_3^{db,l} \q ,
\ea
yielding a new, coarse grained tensor network, tilted by 45$^\circ$. The splitting and contraction of vertices is illustrated in the figures \ref{lattice} and \ref{fig:cvertex} respectively. 

It is possible to implement the TNR method in such a way that the Gau\ss~constraints are explicitly preserved at each coarse graining step \cite{dem,Singh:2009cd}. 
 This is desirable because it preserves this piece of geometrical interpretation in the otherwise very abstract framework of tensor networks. It also increases both speed and stability of the numerical algorithm. 

\begin{figure}
\begin{center}
 \mbox{
      \subfigure[$D_c=16$]{\includegraphics[width=0.45\textwidth]{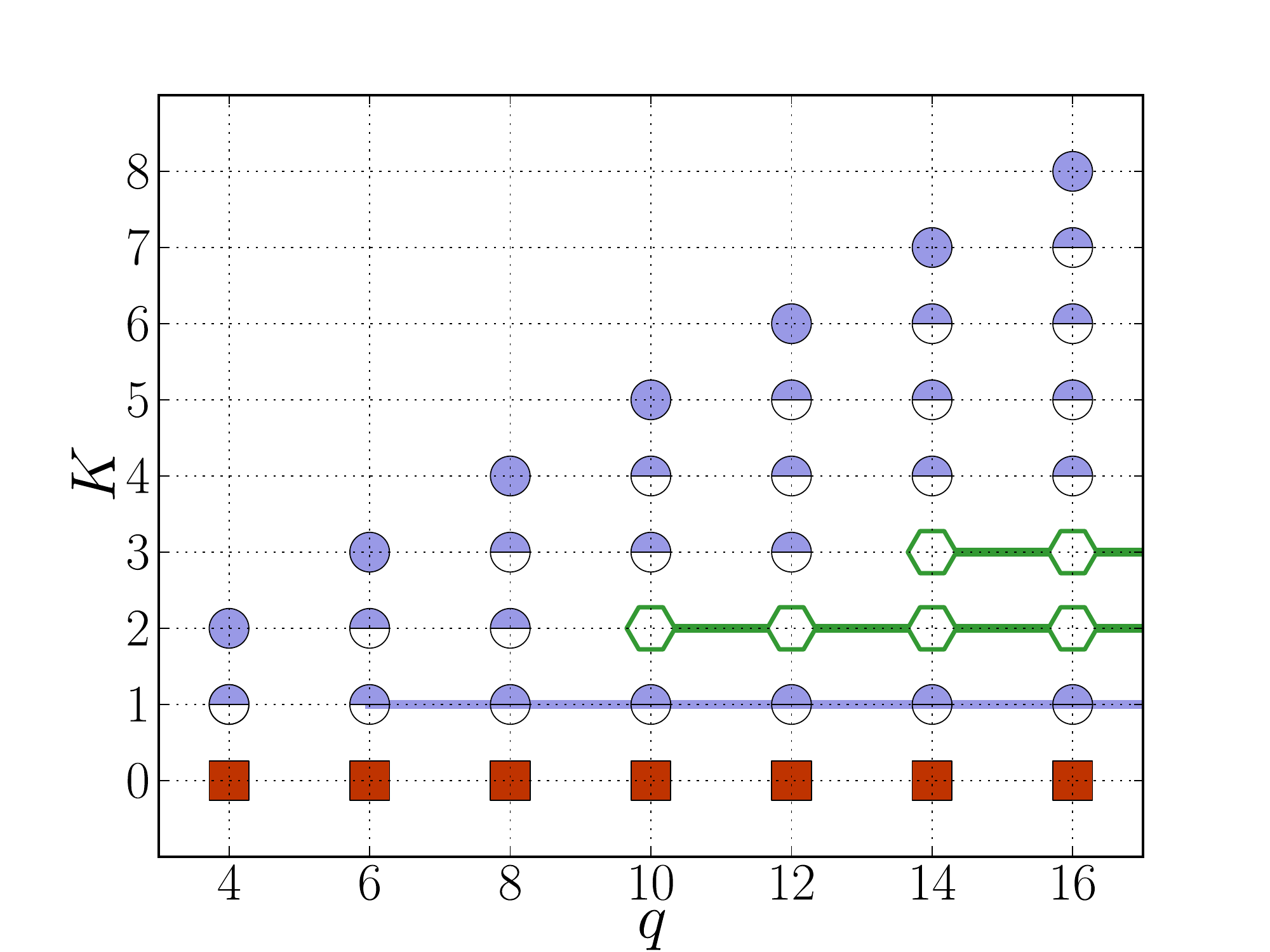}} \quad
      \subfigure[$D_c=25$]{\includegraphics[width=0.45\textwidth]{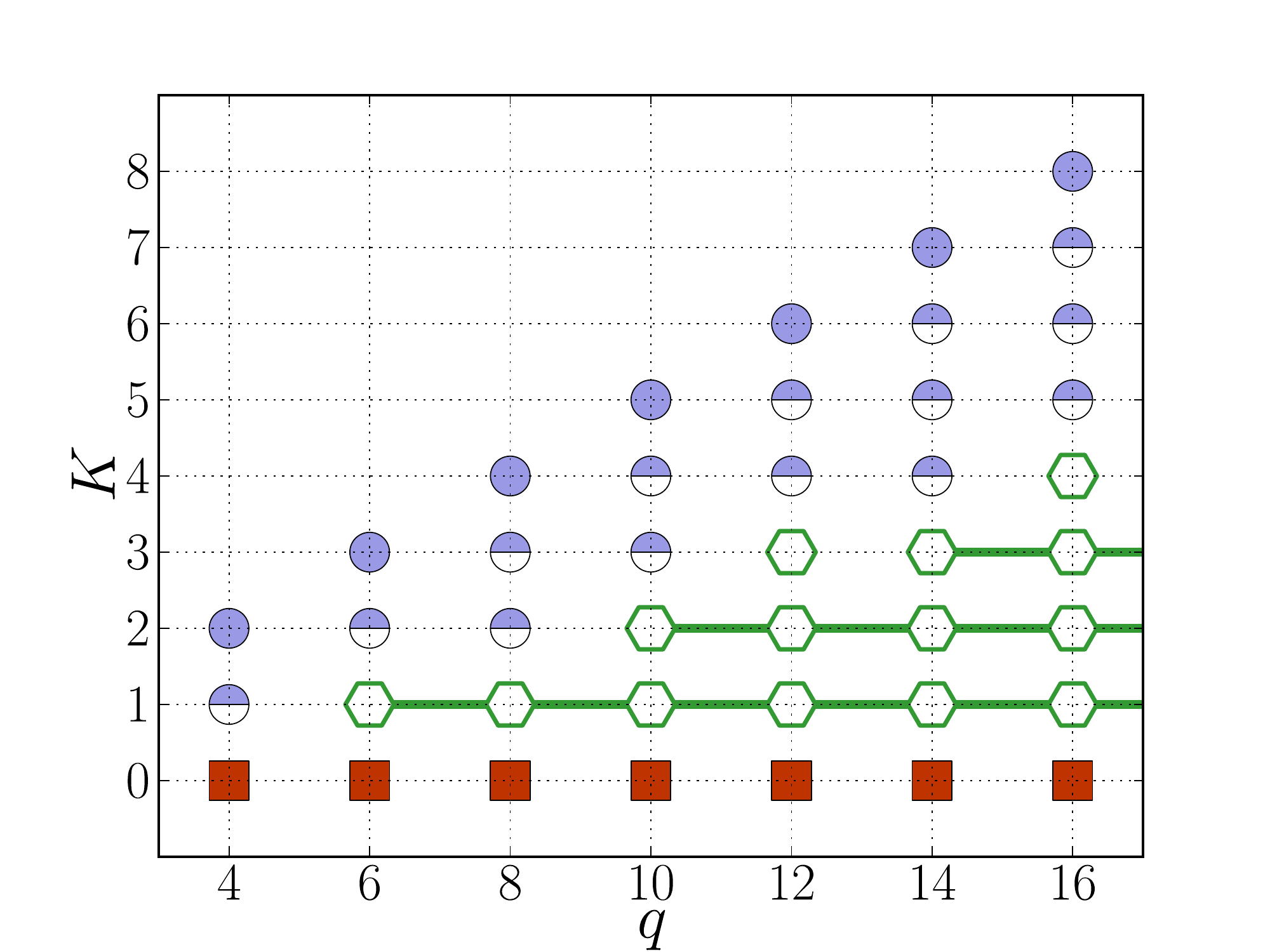}}
     }
\caption{Flow behaviour of different cutoff models, labelled by $K$ and $q$. 
Markers half-filled at the top flow to LTF/LTF$\times$CDL (diagonal models, $K = q/2$), empty hexagonal markers indicate a quasi fixed point or oscillating behaviour. Connected markers illustrate equivalent models. \label{tnr-simus}
\label{tnrkq}}
\end{center}
\end{figure}

Similar to the MK approximation, we also observe quasi fixed points in the tensor network renormalization of cutoff models. Moreover, because of the larger number of parameters compared to MK, the TNR exhibits a richer fixed point structure, including non-isolated fixed points, given e.g.\ by Corner Double Line (CDL) tensors \cite{levin,*Gu:2009dr,dem}. 

In our TNR simulations, the overall results of the MK analysis of cutoff models is confirmed (see figure \ref{tnr-simus}). However, the simulations also show the unique features of the TNR approach: additional CDL fixed point structures for some LTF points and equivalence of models. 

Similar to our MK studies, we also observe some instabilities which lead to oscillating singular values and breaking of the equivalence of models. On the one hand, these effects occur when the models are in the weak phase transition regime and therefore particularly sensitive. On the other hand, numerical instabilities and the rigid imposition of the cutoff lead to asymmetric configurations. Further research is devoted to understanding the instabilities and refine the algorithm accordingly. Thus it may be possible to decide with these refined algorithms whether for instance the Kosterlitz-Thouless transition, which is missed by the MK method, can be observed with the TNR method. 
    
\section{Next steps include refined algorithms and less simplification}

We have argued that accessing and understanding the large-scale structure is crucial for the spin foam approach to quantum gravity. The work presented here proposes a new way of tackling the issue, starting out from simplified, yet feature-rich models and a statistical physics mindset. 
The simplified models are an ideal test bed for developing coarse graining and renormalization methods. Furthermore, studies with increasingly more complicated models will make it possible to put forward and test conjectures about the gravitational spin foam models.

We hope that this line of research will significantly advance our grasp on
the large-scale dynamics encoded in loop quantum gravity and spin foam models. This will be key
to construct improved models, even with the potential to obtain uniqueness results or falsifying the
approach. 

The first steps we considered here opens up numerous opportunities for further research. The observed instabilities in the TNR scheme pose an immediate challenge we would like to tackle. Possible refinements of the algorithm include choosing a symmetric parametrization for the models and measures to ensure symmetry in the framework we use so far. In particular, an adaptive or block-dependent cutoff could be implemented, along with a projection onto symmetric configurations in order to eliminate numerical errors.

Then, in order to make a connection with the mature spin foam models, less simplified models have to be constructed that are still statistically analyzable. This makes it necessary, in particular, to move to non-Abelian groups and, in the TNR scheme, to higher dimensions. A class of models with finite groups that resemble the gravitational spin foam models \cite{eprl} is in the making \cite{bbfw}. Here the main question will be whether a HTF phase, corresponding to a geometrically degenerate phase, will continue to dominate the behavior in the limit of infinite group size, as is the case for standard lattice gauge theories. Current hopes, however,  rely on a flow towards a $BF$ like fixed point \cite{Rovelli:2011fk}. 
For such a fixed point the restoration of diffeomorphism symmetries can also be expected \cite{Bahr:2009qc,*Bahr:2010cq,*Bahr:2011uj}. 

In addition, the application of TNR methods in a canonical formalism is promising \cite{vidalprl,*gulevinwen,*Cirac}. 
 Here, as an improvement of the mean field approach,  the TNR methods would provide a variational ansatz for the physical wave function and an efficient way to compute expectation values of observables, such as the master constraint. These methods can thus help in the master constraint  \cite{Thiemann:2003zv,*Dittrich:2004bp,*Dittrich:2004bn} or uniform discretization program \cite{Campiglia:2006vy,*Gambini:2009ie} to find wave functions minimizing the violation of the master constraint and thus providing an approximation to physical (vacuum) states.

Further interesting questions comprise the generalization to random rather than regular lattices \cite{Markopoulou:2000yy,*Markopoulou:2002ja}, the relationship between degenerate phases and choice of projectors as well as the relationship between the phase structures of fixed lattice and sum over lattices models \cite{Bonzom:2011zz,*Benedetti:2011nn,*Bonzom:2011ev,*Carrozza:2011jn,*Rovelli:2010qx}. 

The plethora of relevant research opportunities encourages us to advocate the statistical approach to understanding the large-scale structure of spin foams. We see great benefits in this connection between quantum gravity and statistical / condensed matter physics and hope to contribute to cross-fertilization in both communities.

\ack We would like to thank Mercedes Martin-Benito for collaboration on \cite{dem}.

\bibliographystyle{iopart-num}
\bibliography{proceed}

 \end{document}